\documentclass[]{raa}            
\usepackage{graphicx,times}
\usepackage{natbib}

\providecommand{\bm}[1]{\mbox{\boldmath$#1$\unboldmath}}

\begin{document}

   \title{Metallicity calibration for solar type stars based on red spectra
}

 \volnopage{ {\bf 2009} Vol.\ {\bf 9} No. {\bf XX}, 000--000}
   \setcounter{page}{1}

   \author{J. K. Zhao
   \and G. Zhao
   \and Y. Q. Chen
   \and A. L. Luo
   }

   \institute{Key Laboratory of Optical Astronomy, National Astronomical Observatories, Chinese Academy
of Sciences, Beijing 100012, China; {\it gzhao@bao.ac.cn}\\
        \vs \no
   {\small Received [year] [month] [day]; accepted [year] [month] [day] }
}

\abstract{ Based on high resolution and high signal-to-noise ratio (S/N) spectra
analysis of 90 solar type stars, we have established several new
metallicity calibrations in $T\rm{_{eff}}$ range [5600, 6500] K based on red spectra with the
wavelength range of 560-880\,nm. The new
metallicity calibrations are applied to determine the metallicity
of solar analogs selected from SDSS spectra. There is a good
consistent result with the adopted value presented in SDSS-DR7 and
a small scatter of 0.26 dex for stars with S/N $>$ 50 is obtained.
This study provides a new reliable way to derive the metallicity for solar-like stars with low resolution spectra. In particular, our
calibrations are useful for finding metal-rich stars, which are
missing in SSPP.
}

   \authorrunning{J. K. Zhao, G. Zhao, Y. Q. Chen, A. L. Luo }            
   \titlerunning{Metallicity calibration for solar type stars }  
   \maketitle


%
%
\section{Introduction}           
\label{sect:intro}

The stellar spectroscopic survey with the Large Area sky
Multi-Object fiber Spectroscopic Telescope (LAMOST) will provide a huge
amount of data, which can be used for the study of chemical
and kinematical evolution of our Galaxy. In this respect, stellar
metallicity and radial velocity, being two main parameters, can be
derived from spectra. The determination of radial velocity is generally easier mainly by using either cross-correlation of the template
spectra or Doppler shift through line calibration. The
consistency is usually quite good depending on the quality of the spectra. The
metallicity estimation from stellar spectra is
based on various methods as shown in Lee et al.
(2008a) (hereafter Lee08). However, for solar type stars, these values can be underestimated by up to 0.5 dex in the previous
version of SSPP (SEGUE Stellar Parameter Pipeline; Lee et al. 2008a). The current version of SSPP has made great
improvement, reaching about 0.1 dex in the underestimation (Lee et al. 2008b). From
Fig A1 in Bond et al. (2009), the largest difference between the SDSS (Sloan Digital Sky Survey; York et al. 2001)
spectroscopic metallicity values with DR6 (Data Release 6; 	
	Adelman-McCarthy et al. 2008) and DR7 (Data Release 7; Abazajian et al. 2009) is shown for solar type stars, so it is worth the effort to do more research about deriving the metallicty of those stars.


In this work, we attempt to establish a new metallicity
calibration for low resolution solar type stars based on the
result from high resolution and high signal-to-noise ratio spectral
analysis performed by Chen et al. (2000, hereafter Chen00). In comparison
with the methods presented in Lee08, this work has some
advantages. Firstly, the calibration is based on the real stellar
(empirical) spectra and their metallicity is derived from fine
analysis of high resolution and high S/N spectra.
Secondly, we have used the red spectral coverage of 560-880\,nm
but most of the methods in Lee08 are based on blue spectra
with $\lambda <600 $\,nm. As is well known, the advantage of the red
spectra is easier to define continuum, which is not possible for
blue spectra due to the heavy line blanketing at low resolution
observations. In view of this advantage, we adopted the equivalent
widths (EW) of individual lines in the calibration instead of the line
index. As for line index, there might be different definition for different authors£¬ while EW is a fixed value. For example, CaII K line, there are K6, K12 and K18 among its definition. If we can define the continuum well, the EW is better than line index. Thirdly, we have adopted only
Fe lines for metallicity calibration and avoid contributions from
other elements, which do not exactly trace iron evolution at
different times and different nucleosynthesis sites. In Lee08, the
wavelength ranges of templates match include all lines from
different elements. The KP line index is also an indicator of [Ca/H].
Although these weak Fe lines are undetectable in metal-poor stars
because of the noise, we can recognize them in solar type stars
where the S/N is higher than 20.  Moreover, the EW of weak lines is more sensitive to abundance
than that of strong lines. For high resolution spectra analysis, strong lines, e.g. Na 5895$\rm \AA$ and Na 5890$\rm \AA$ are saturated and in the growth curve the increasing EWs do not give higher abundance. In general, the EW of strong lines does not change a lot with the degradation of resolution. In view of this, it is not optimal to establish relation between abundance and EW in combination with colors using very strong line. Finally,
the calibration is internally consistent, while Lee08 adopted the
average of different values from various methods.

In Section 2, a description of the data for the calibration is presented. The template matching analysis
is described in Section 3. The detailed analysis procedure to get the calibration formula is given in Section 4. The application of the calibration to SDSS spectra is illustrated in Section 5. Finally, the conclusion is given in
Section 6.


\section{Data}
The real spectra are taken from Chen00 which has a resolving power of
37\,000 and S/N of 150-300 obtained with the Coud$\acute{e}$
Echelle Spectrograph mounted on the 2.16 m telescope of the National
Astronomical Observatories (Zhao $\&$ Li 2001). The sample has $T\rm{_{eff}}$, log $g$  and [Fe/H] distributions as shown in Fig. 1. It is clear that $T\rm{_{eff}}$  ranges between [5600, 6500],  log $g$ is in [3.98,
4.43] and  [Fe/H] is in [-1.04, 0.06]. Moreover, the range
of b-y is within [0.28, 0.43], B-V is within [0.40, 0.67], V-I is within [0.46, 0.73] and V-K is within [0.9, 1.65]. We
convolved the normalized spectra to low resolution of 2000 with
a Gaussian Function. In addition, the spectra were rebinned to 1.5$\rm{\AA}/pix$ after smoothed to R$\sim$2000.
\begin{figure}
   \begin{center}
  \includegraphics[width=100mm]{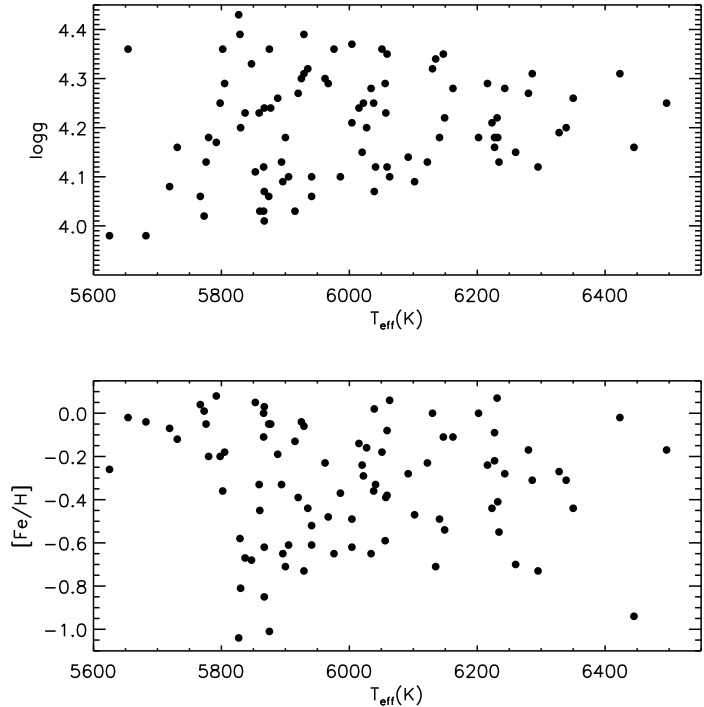}
   \end{center}
   \caption{Stellar parameter distribution of the sample in Chen00  }
\end{figure}
\begin{table}
\centering
\begin{minipage}{80mm}
\caption{[Fe/H] results of template match with different
wavelength ranges} \label{atmos_parameter1}
\begin{tabular}{ccc}\hline
wavelength range(nm)&mean deviation&scatter\\
\hline
  \noalign{}
 570-653&     0.448&       0.175  \\
  \hline
 570-684&   0.226&         0.275  \\
  \hline
 570-700&   0.016&         0.230 \\
  \hline
 640-670&  0.470&          0.280 \\
  \hline
 651-662&  0.599&          0.390 \\
  \hline
 690-713&  -0.07&          0.590 \\
  \hline
 735-756&    0.350&        0.290 \\
 \hline
 772-810&    -0.100&        0.260 \\
  \hline
  \noalign{}
  \end{tabular}
  \end{minipage}
 \end{table}

 \section{The template match analysis}
Following one method of Lee08, we have performed the template
spectra match (also see Allende et al 2006, Re Fiorentin
et al 2007, Zwitter et al. 2005) for the normalized low
resolution spectra of 90 stars and derived the stellar
temperature, gravity and metallicity. In this method, we generate
a library of low resolution theoretical spectra by using the SYNTH
program based on Kurucz New ODF (Castelli $\&$ Kurucz 2003)
atmospheric models.  The atmosphere models are under the assumption of local thermodynamic equilibrium (LTE). The mixing-length is adopted to be l$/$Hp = 1.25 and  microturbulence is 1.5 km s$^{-1}$. The line list, including the atoms and molecules, are all from Kurucz (1993). The molecular species include CH, CN, OH, and TiO. Solar abundances are from Asplund (2005). As for those grids, $T\rm{_{eff}}$ covers the range [3500-9750]K with 250K interval; log $g$ is within [1.0, 5.0] dex with 0.5 dex interval; [Fe/H] is within [-4.0, -3.0] dex with 0.25 dex interval and 0.1 dex interval in [-3.0,+0.5] dex. The minimum distance method is applied to
obtain the parameters by interpolation among several of the closest
theoretical spectra with the observed one.
\begin{figure}
   \vspace{1mm}
   \center
   \hspace{1mm}\includegraphics[height=80mm,width=80mm]{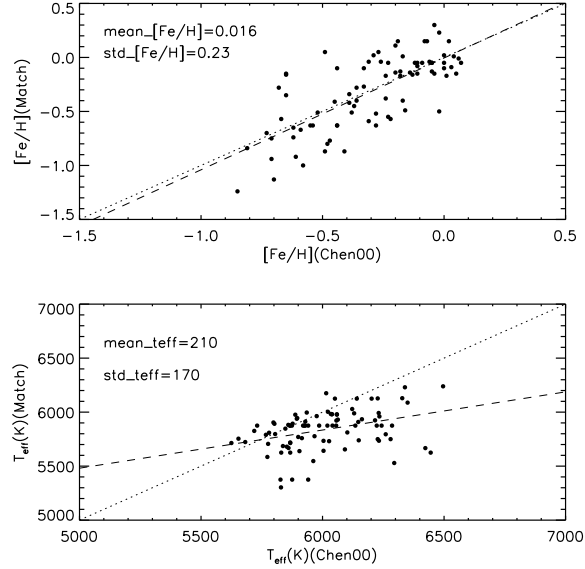}
     \caption{[Fe/H] and $T\rm{_{eff}}$ comparison between those of Chen00 and the results from the template match with spectral range of 570nm-750nm  }
\end{figure}

We have adopted different wavelength ranges in the matching
procedure and obtained different results as shown in Table 1. As
compared with the `standard' values presented in Chen00 paper,
we have found that the spectral range of 570-700\,nm
is the best choice with a mean deviation of 0.016 dex and
scatter of 0.23 dex in [Fe/H]. The comparison of temperature and
metallicity of the 570-700\,nm spectral range is shown in Fig. 2. It is clear
that the temperature estimation has systematical deviation with
a lower value in the present work.  Since the high resolution spectra in Chen00 are obtained with echelle spectrograph. The order is not wide enough to include the whole H alpha line region and the normalization is implemented order by order. Thus, the continuum is not very reliable in H alpha region, which might be the reason of systematical deviation in temperature estimation.
The metallicity is quite consistent with an rms of 0.23 dex.

\section{[Fe/H] vs. EW of the FeI line - An empirical calibration}
Although the template match method can be used to obtain accurate stellar
parameters, it may give different results with different wavelength
ranges. Hence, we will determine stellar metallicity
based on the strength of some FeI lines.

\begin{figure}
   \vspace{1mm}
\center
  \hspace{1mm}\includegraphics{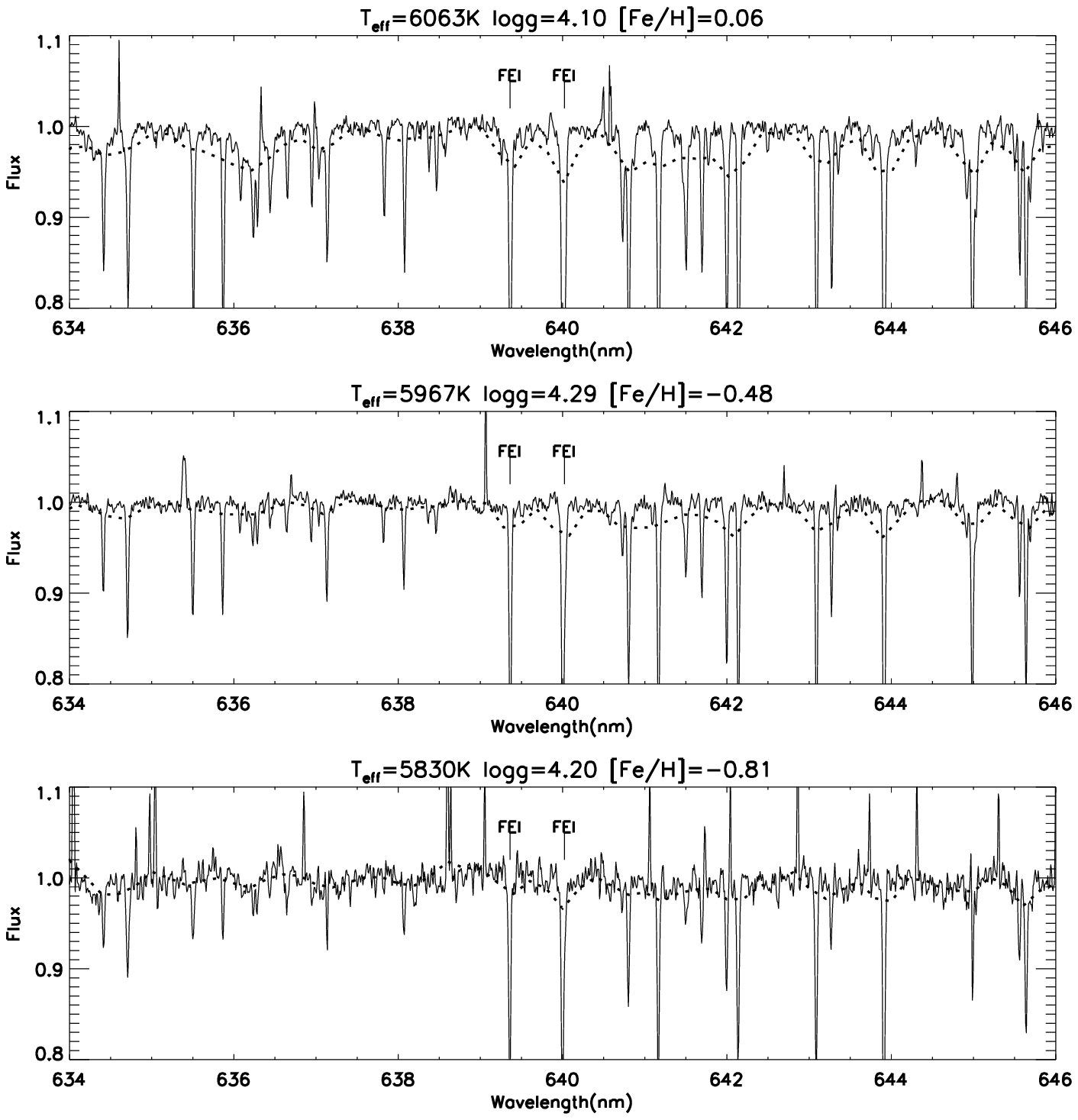}

   \caption{A portion of the spectra of HD94280, HD100446 and HD184601. Solid line is the high resolution spectra while thick dotted line is the low resolution spectra. The two FeI lines
           are those that meet the stated requirements.   }
\end{figure}

\begin{table}
\centering
\begin{minipage}{80mm}
\caption{The definition of FeI lines.}
\label{atmos_parameter}
\begin{tabular}{cccc}\hline
Left&Right&Center&Element\\
\hline
  \noalign{}
   6060.917 &    6069.954 &    6065.492 &  FeI \\
   \hline
   6217.769&     6221.788 &    6219.292 &  FeI\\
   \hline
   6391.147&     6395.953 &    6393.605 &  FeI \\
   \hline
    6398.256&    6402.209&     6400.232& FeI\\
   \hline
   6675.000&     6682.326&    6678.256&  FeI \\
   \hline
  \noalign{}
  \end{tabular}
  \end{minipage}
   \end{table}
\subsection{The spectral lines selection}
What we want is to choose some Fe lines that are not heavily blended
and have better profiles in the low resolution (R$\sim$2000) spectra. At the
beginning, we draw the original spectrum, then overplot the low
resolution spectrum on it. After checking the lines one by one, we
select five FeI lines as our indices of metallicity.
The spectral lines for three stars with different
metallicity values are given in Fig. 3: HD94280 has [Fe/H] of 0.06; HD100446 has -0.48;
HD184601 has -0.81. In Fig. 3, the solid line is the high
resolution spectrum while the thick dotted line is that of the low
resolution one. In this segment of the spectrum, only two FeI lines meet
our requirements since they are detectable; they have good shapes
and are not seriously blended in the spectra with a resolution of R$\sim$2000.
From the top to the bottom of Fig. 3, it is obvious that the strength
of FeI lines decreases. Also, for stars with [Fe/H] $>$ -0.8, the
two lines can be identified. Table 2 presents the definition
of five FeI line indices used for our metallicity calibration.
There are three parts for each line including red, center and
blue spectra. The EW of each line can be measured by using a direct integration method.

\begin{figure}
   \vspace{1mm}
   \center
   \hspace{1mm}\includegraphics{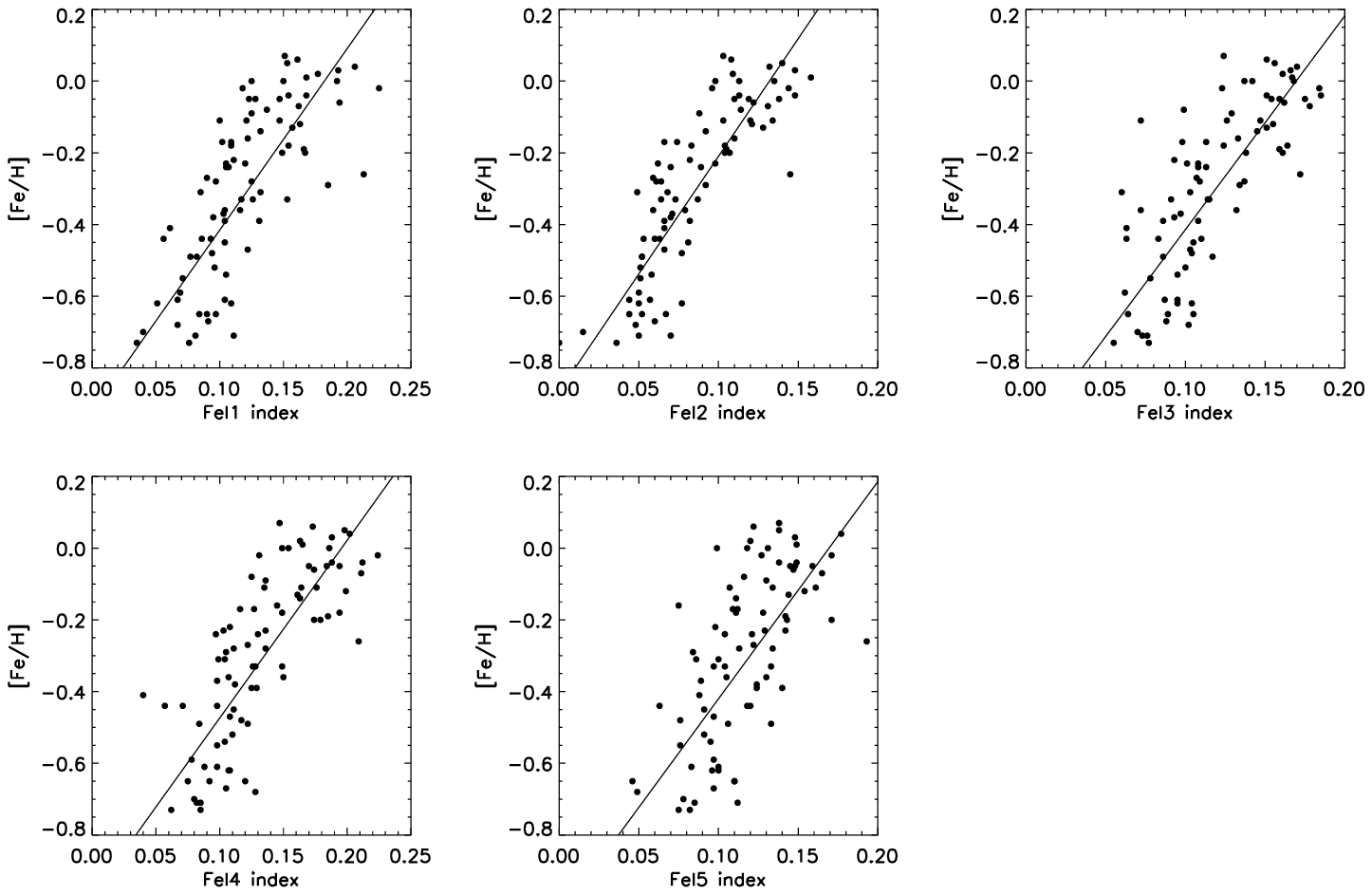}

   \caption{The relation between [Fe/H] and EW for five FeI lines for the stars with [Fe/H] $>$ -0.8 }
\end{figure}
\begin{table}
\centering
\begin{minipage}{80mm}
\caption{The coefficient and $\sigma$ of the fitting between
 [Fe/H] and EW for each FeI line}
\label{atmos_parameter1}
\begin{tabular}{cccc}\hline
Lines& a & b & $\sigma$ \\
\hline
   \noalign{}
 FeI1&      -0.922 &      5.063   &    0.170 \\
  \hline
   FeI2&     -0.866&       6.559   &    0.143  \\
  \hline
   FeI3&    -1.010  &     5.960  &     0.167 \\
  \hline
   FeI4&     -0.970  &     4.958   &    0.163 \\
  \hline
   FeI5&    -1.026 &      6.056  &     0.204 \\
    \hline
  \noalign{}
  \end{tabular}
  \end{minipage}
 \end{table}

 \begin{table}
\centering
\begin{minipage}{80mm}
\caption{The coefficient and $\sigma$ of [Fe/H] calibration
based on the EW of five FeI lines and temperature }
\label{atmos_parameter1}
\begin{tabular}{ccccc}\hline
Lines& a & b &c& $\sigma$ \\
\hline
   \noalign{}
  FeI1&       1.731   &    5.929   &   -3.281    &   0.150 \\
  \hline
   FeI2&      2.138   &    7.726    &  -3.692    &   0.112\\
  \hline
   FeI3&     2.282    &   7.386   &   -4.116    &   0.136\\
  \hline
   FeI4&    2.793  &     6.381    &  -4.702    &   0.122 \\
  \hline
   FeI5&   0.974   &    6.949   &   -2.505    &   0.194 \\
    \hline
  \noalign{}
  \end{tabular}
  \end{minipage}
 \end{table}

\subsection{The metallicity calibration based on EWs}
Since our spectra are normalized and there are no spectra with
flux calibration, it is difficult to derive reliable temperature measurements.
In order to improve the metallicity determination, we
resort to using the strengths of iron lines and establish a
calibration between metallicity and EWs of iron lines. The EWs are derived with Equation 1.

\begin{eqnarray}
\rm{EW}&=&\int_{\lambda1}^{\lambda2}{\frac{f_{c}-f_{\lambda}}{f_{c}}d_{\lambda}}
\end{eqnarray}

In Equation1, the f$_{\lambda}$ represents the flux of wavelength $\lambda$, while f$_{c}$ means the continuum of
wavelength $\lambda$.
 It is shown in Fig. 4 that there is a good correlation between [Fe/H] and EWs for the
five FeI lines for stars whose [Fe/H] $>$ -0.8. The calibration (Equation 2) for each line is
shown in Table 3.
\begin{eqnarray}
\rm[Fe/H]&=&a+b*\rm{EW(FeI)}
\end{eqnarray}
As seen in Fig. 4, FeI2 and FeI4 show the best result with the
lowest scatter in the relation. To show the temperature effect of
EW, we divide the temperature range into three parts. The first
part is the range of [5265, 5900]; the second part is [5900,
6200] and the last part is [6200, 6496]. In Fig. 5, the stars
in first part are given with the sign of the dot and asterisks represent
the stars in second part, while the stars in last part are
plotted with diamonds. From Fig. 5, it is clear that
the relation between EW and [Fe/H] changes with temperature. FeI2
and FeI4 have lower scatter than other FeI lines and this
may be due to the lower sensitivity of line strength with
temperature. In order to understand this issue, we added the
temperature term in the calibration in Fig. 6. The coefficients
and scatter of each line are given in Table 4. It is obvious that
the $\sigma$ becomes small for the calibration of each line after
considering the effect of temperature. Since the temperature is
usually unknown in the spectra analysis, it may be good to replace
the temperature term with the color index. Thus, we collected (b-y),
(B-V), (V-I) and (V-K) and performed a similar
calibration (Equation 3). 
log $g$ also will bring more or less uncertainty on metallicity determination. However, the gravity range
considered in the calibration sample is narrow, hence its effect is very little and could be ignored.

Table 5 is the coefficients and scatter of the calibration of [Fe/H] through EW
and B-V.
\begin{table}
\centering
\begin{minipage}{80mm}
\caption{The coefficient, $\sigma$ and the EW range of the fitting between
   [Fe/H] and the EW plus B-V for each FeI line}
\begin{tabular}{cccccc}\hline
Lines & a & b & c & $\sigma$&EW range \\
\hline
  \noalign{}
   FeI1&     -0.790&       5.351&      -0.315&       0.169& 0.025$\sim$0.225\\
  \hline
   FeI2&     -0.290&       8.144&      -1.342&       0.132&0.006$\sim$0.158\\
  \hline
   FeI3&     -0.555&       7.352&      -1.167&       0.160&0.044$\sim$0.185\\
  \hline
   FeI4&     -0.321&       6.622&      -1.643&       0.151&0.029$\sim$0.224\\
  \hline
   FeI5&     -1.200&       5.497&      0.453&       0.202&0.046$\sim$0.193\\
      \hline
  \noalign{}
  \end{tabular}
  \end{minipage}
 \end{table}

\begin{eqnarray}
\rm[Fe/H]&=&a+b*\rm{EW(FeI)}+c*(B-V)
\end{eqnarray}

\begin{figure}
   \vspace{1mm}
   \center
   \hspace{1mm}\includegraphics{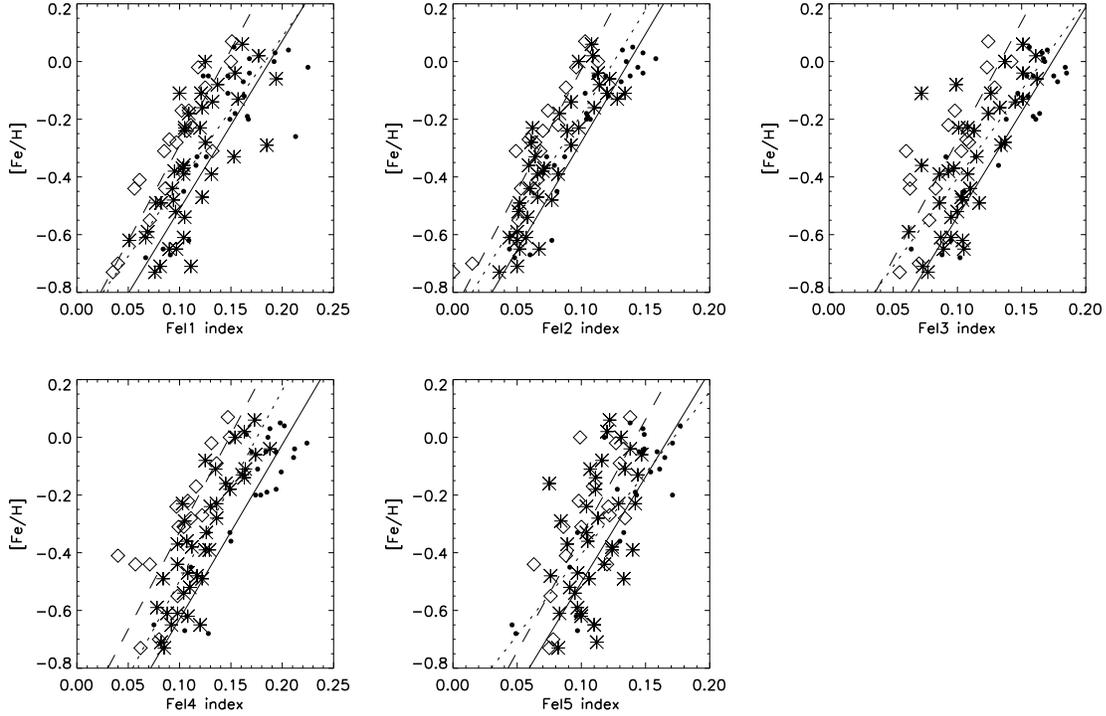}

   \caption{The relation between the metallicity and EW of FeI lines in different $T\rm{_{eff}}$ ranges.
    Dots represent stars in [5265, 5900]K; asterisks are those in [5900, 6200]K; diamonds are those in [6200, 6500]K.}
\end{figure}

\begin{figure}
   \vspace{1mm}
   \center
   \hspace{1mm}\includegraphics{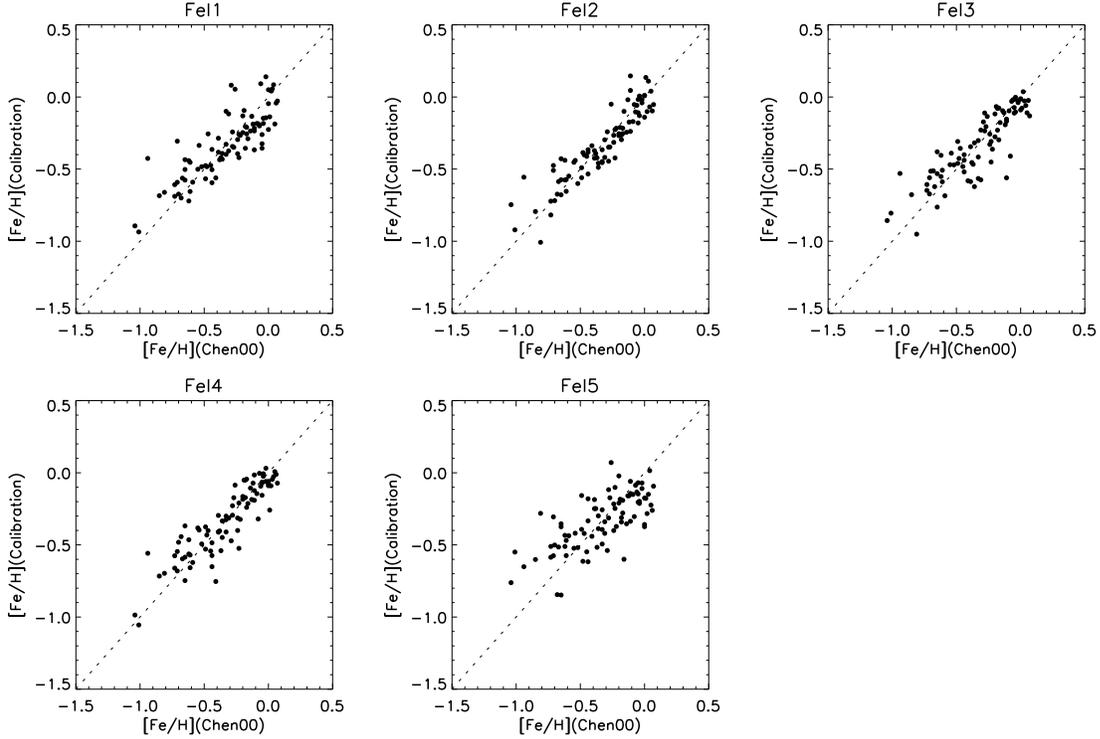}
   \caption{The comparison between [Fe/H] in Chen00 and that of calibration from each FeI line and the temperature }
\end{figure}

\begin{figure}
   \vspace{1mm}
\center
   \hspace{1mm}\includegraphics[width=84mm]{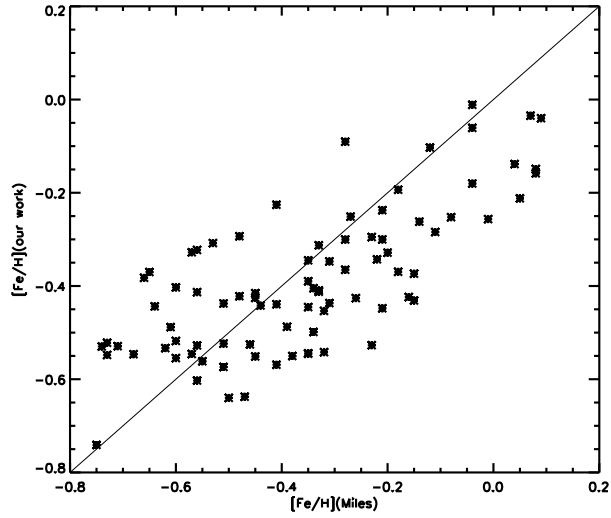}
   \caption{The application of our [Fe/H] calibration in Miles spectra library. The x axis is [Fe/H] from the Miles catalog and the y axis is [Fe/H] obtained by our calibration. The solid line is x=y }
\end{figure}

\begin{figure}
   \vspace{1mm}
   \center
   \hspace{1mm}\includegraphics[height=80mm,width=80mm]{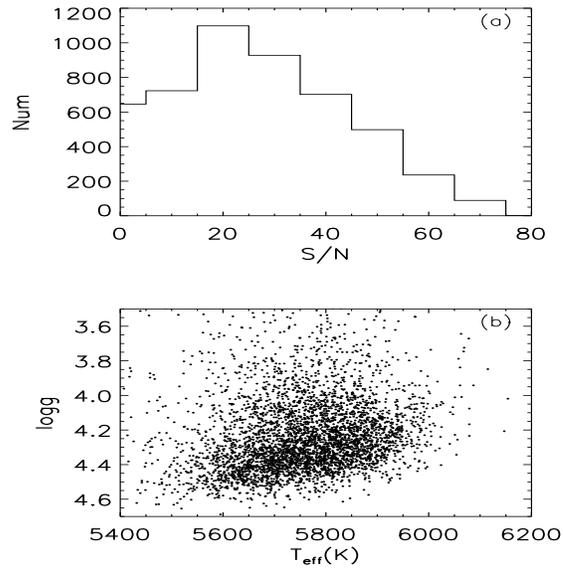}
   \caption{Plots of (a) number vs. S/N (b) log $g$ vs.  $T\rm{_{eff}}$}
\end{figure}

\begin{figure}
   \vspace{1mm}
\center
   \hspace{1mm}\includegraphics{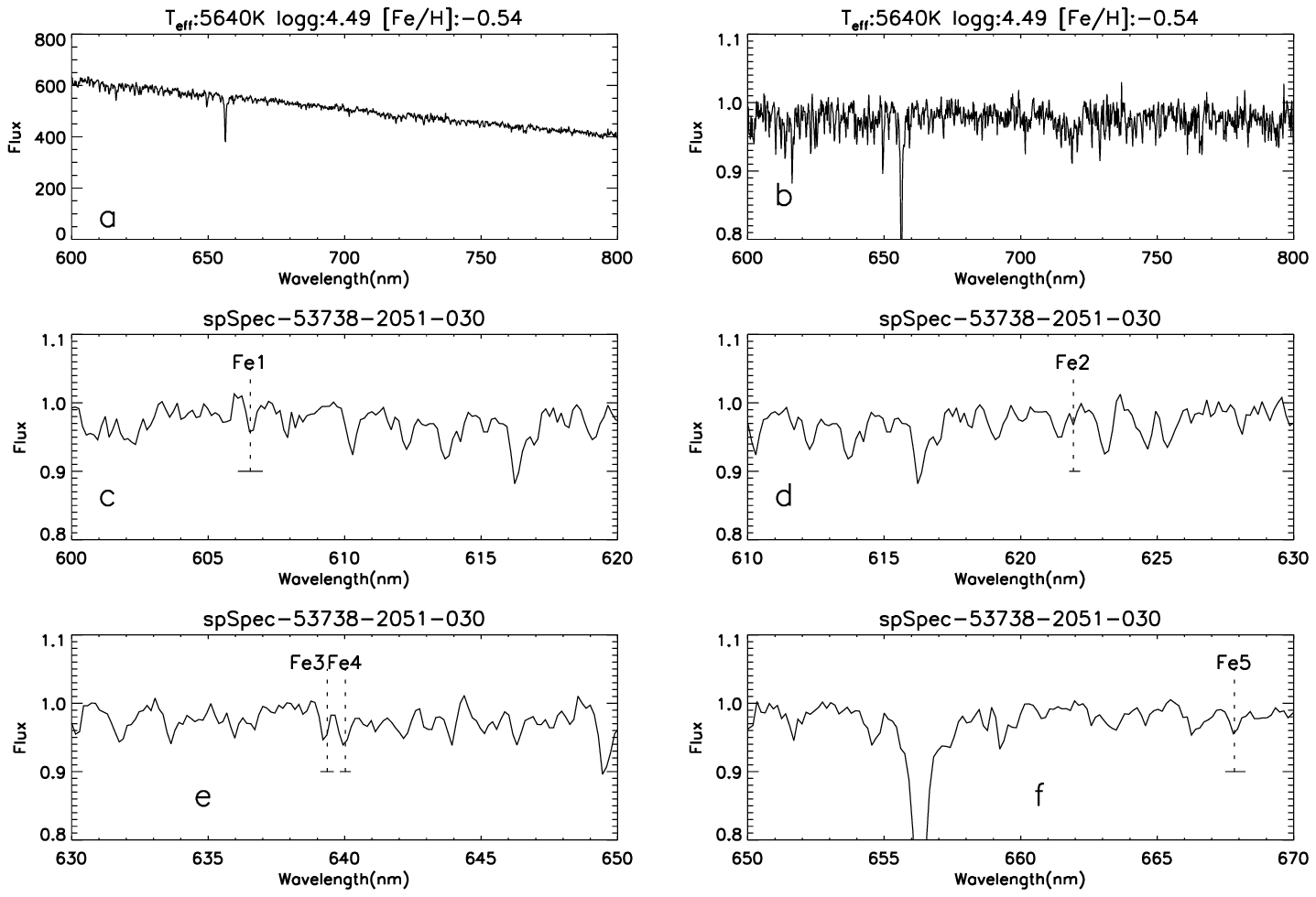}
   \caption{The spectra of 53738-2051-030. `a' is the original spectrum while `b' is normalized spectrum. `c',`d',`e',`f' and `g' present the spectral lines in our calibration    }
\end{figure}

\subsection{Calibration in Miles spectra library}
 To make an external calibration, we selected 107 spectra from the Miles spectral library (S\'{a}nchez-Bl\'{a}zquez et al. 2006; Cenarro et al. 2007) which meet the following conditions:  -0.8 $\leq$ [Fe/H] $\leq$ 0.5; 4.0 $\leq$ log $g$ $\leq$ 4.5; 5600 $\leq$ $T\rm{_{eff}}$ $\leq$ 6500. Thus, it is available to estimate the metallicity of these 107 spectra using above calibrations. The resolution of Miles spectra is about 2.3$\rm \AA$ and the wavelength has already been calibrated with radial velocity. First, we do normalization for these 107 spectra. The continuum is determined by iteratively smoothing with a Gaussian profile, and then clipping off points that lie beyond 1 $\sigma$ or 4 $\sigma$ above the smooth curve. Second, the EWs of five FeI lines are measured. B-V of those 107 stars are taken from the literature identified by the Simbad Astronomical Database (Genova 2006). Finally, the metallicities are derived by EW of FeI4 line using equation 3 (the coefficients is shown in Table 5).  Fig. 7 is [Fe/H] comparison between our results and those from Miles library. The mean error is about 0.08 dex, and the scatter is about 0.21 dex. [Fe/H] of Miles library is obtained by the compilation from the literature.  This suggests that our metallicity is basically consistent with that of other work.


\section{Application of these calibrations}
In order to check the accuracy of [Fe/H] calibration from the
EW of five FeI lines, we implement this calibration to determine
[Fe/H] for solar-like stars with SDSS spectra. The selection
limitation is as follows: $0.4<(g-r)_0<0.5$, $0.10<(r-i)_0<0.14$,
$0.02<(i-z)_0<0.06$, and $g_0<20$. The above color ranges come from the
color of the Sun and its error bars. Based on this limitation, 4356
stars are extracted from the SDSS DR7 archive. Fig. 8 presents the
information of this sample, from which we can see the peak of S/N is 20;
the effective temperature of most stars is located in the range of [5500, 6000]K;
 log $g$ is in [4.0, 4.5]. By transforming equation of Bilir et al.
(2005), the g-r range of Chen00 is [0.197, 0.501], so it is
available to estimate the metallicity of these solar-like stars
using Equation 2 and Equation 3.

\subsection{Preprocess}
Our first preprocessing procedures mainly include radial
velocity correction and normalization. The value of radial velocity comes from the FIT head of each
spectra. The pseudocontinuum is determined with the same method illustrated in Sec. 4.3. Although the method of pseudocontinuum determination is different with that in Chen00, it is good enough for solar type stars in the red spectral region. Since the SDSS spectra are relative flux calibrated, the pseudocontinuum in red region is easier represented by a lower order polynomial. Thus, iteratively
smoothing with a Gaussian profile to original spectrum will get a good continuum shape. After normalization, then the EW of the lines can be
measured by the direct integration method. Fig. 9 is an example of the SDSS
spectrum. `a' is the original spectrum while `b' is the normalized
spectrum. `c', `d', `e' and `f' present the FeI lines in our
calibration. It is clear that these FeI lines are detectable and show a
good profile in the SDSS spectra.
\begin{figure}
\vspace{1mm}
   \center
   \hspace{1mm}\includegraphics[width=130mm,height=130mm]{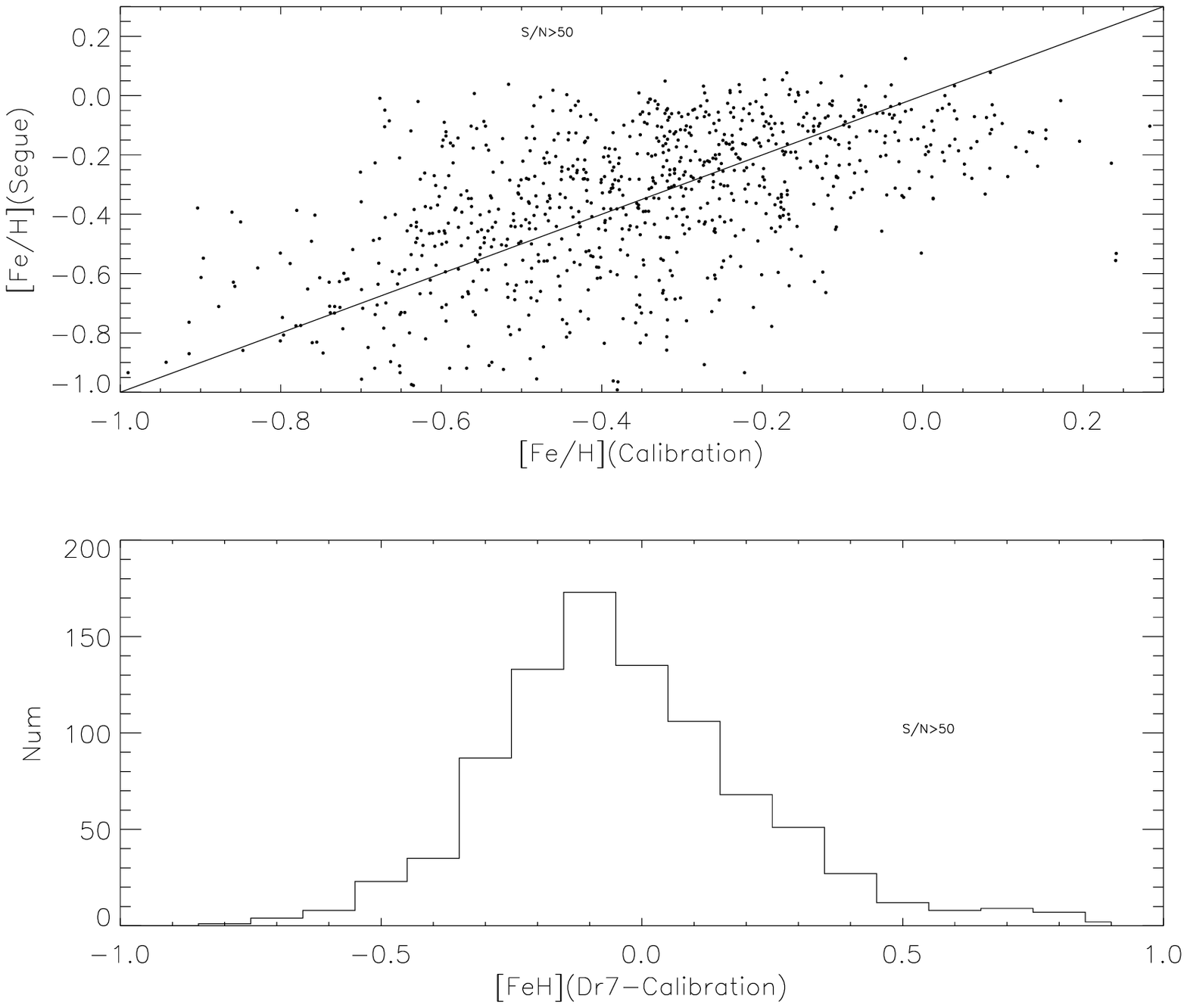}
   \caption{[Fe/H] comparison of the sdss solar-like stars between those from SSPP and those derived from the calibration using five FeI lines  }
\end{figure}

\subsection{Calibration}
After obtaining the EW of five FeI lines, [Fe/H] can be
determined by our calibration formula. We do some comparison
between [Fe/H] from our calibration and
those from SSPP. Equation 4 is the calibration formula based on the FeI4 line. B-V can be obtained from g-r
transformation (Bilir et al. 2005). Fig. 10 is  [Fe/H] comparison between those from SSPP and the results from Equation 4.  The comparison of stars with S/N $>$ 50 is shown in the top panel and the
difference distribution is given in bottom panel. It is
clear that our calibration from Equation 4 is very consistent with [Fe/H]
of SSPP for those with S/N $>$ 50. The mean difference is about 0.018 dex and the scatter is around 0.26 dex.

\begin{eqnarray}
\rm[Fe/H]=-0.321+6.622*\rm{EW(FeI4)}-1.643*(B-V)
\end{eqnarray}

Moreover, we extract 51 stars which meet these conditions:
[Fe/H]$\geq$0 from our result; [Fe/H]$<0$ from SSPP; S/N$>50$. So
these 51 stars are metal rich stars if our result is right. As for these 51 stars, the temperature range is  about [5650, 5865]K and the gravity range is in [4.3, 4.5] dex. To
check the reliability of our result, we make a comparison between the
strength of some spectral lines in these stars and those in the Sun
since $T\rm{_{eff}}$ and log $g$ of these stars are very close to those of
the Sun. If the strength is stronger than that of the Sun then
this star can be regarded as a metal rich star and its [Fe/H] is
larger than 0. Since the NaI and CaII line are strong in the red band,
these lines, as well as two FeI lines, are selected for comparison. There
are three cases:  one is that the strengths of these lines are all
stronger than those of the Sun (see Fig. 11); one is that the
strengths of these lines are all close to those of the Sun (See
Fig. 12); the others are taken as the third case (See Fig. 13). After
comparison, there are 33 stars in first case, 10 stars in second
case and 8 stars in third case. In view of this, the metal rich
stars account for 84\% of these 51 stars. So our calibration also
provides a reliable way to identify metal rich stars.

\begin{figure}
   \vspace{1mm}
   \center
   \hspace{1mm}\includegraphics[width=84mm]{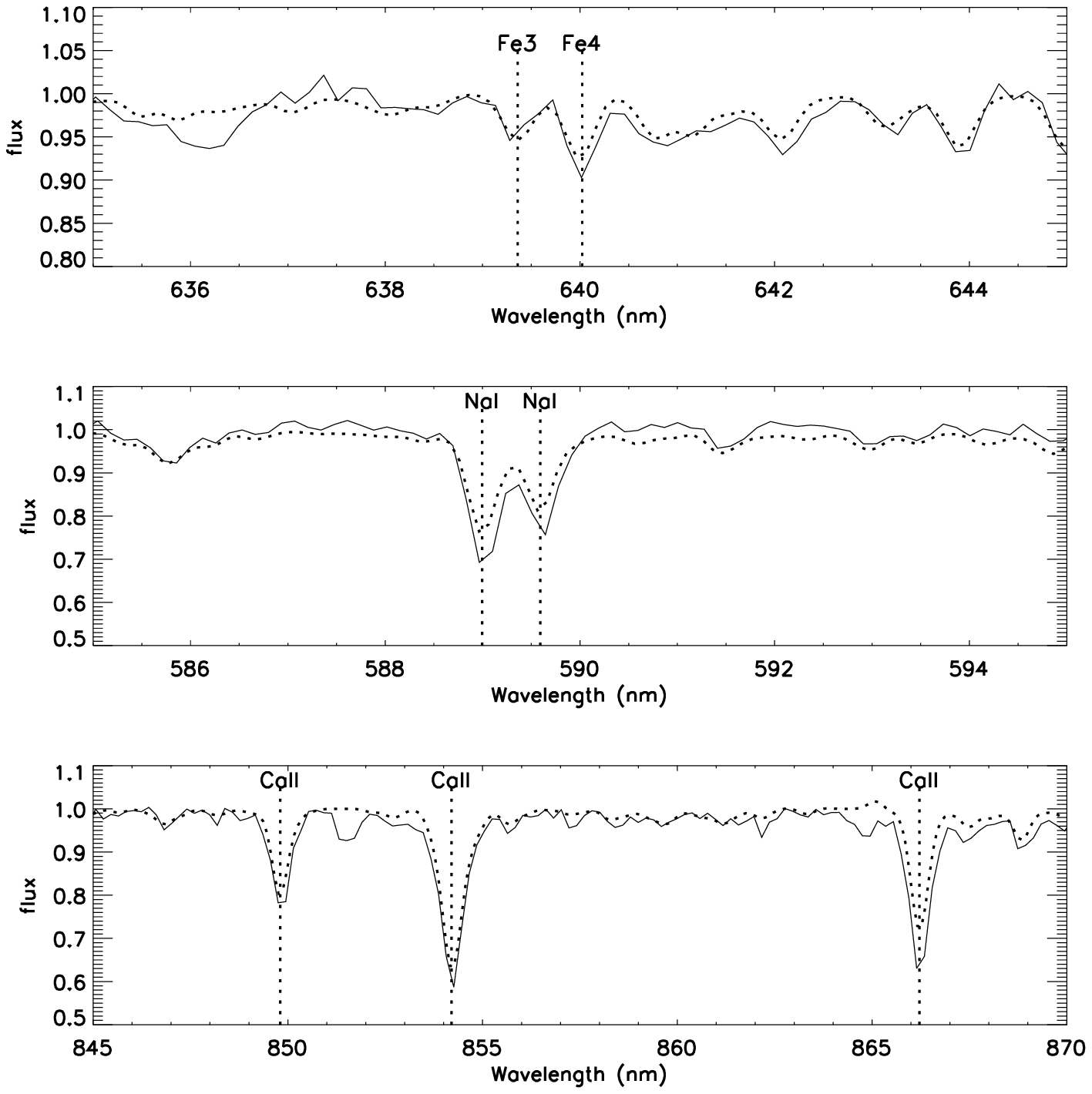}
   \caption{The comparison of the strength of spectral lines. The solid line is the object spectrum and the dotted line is the solar spectrum}
\end{figure}
\begin{figure}
   \vspace{1mm}
   \center
   \hspace{1mm}\includegraphics[width=84mm]{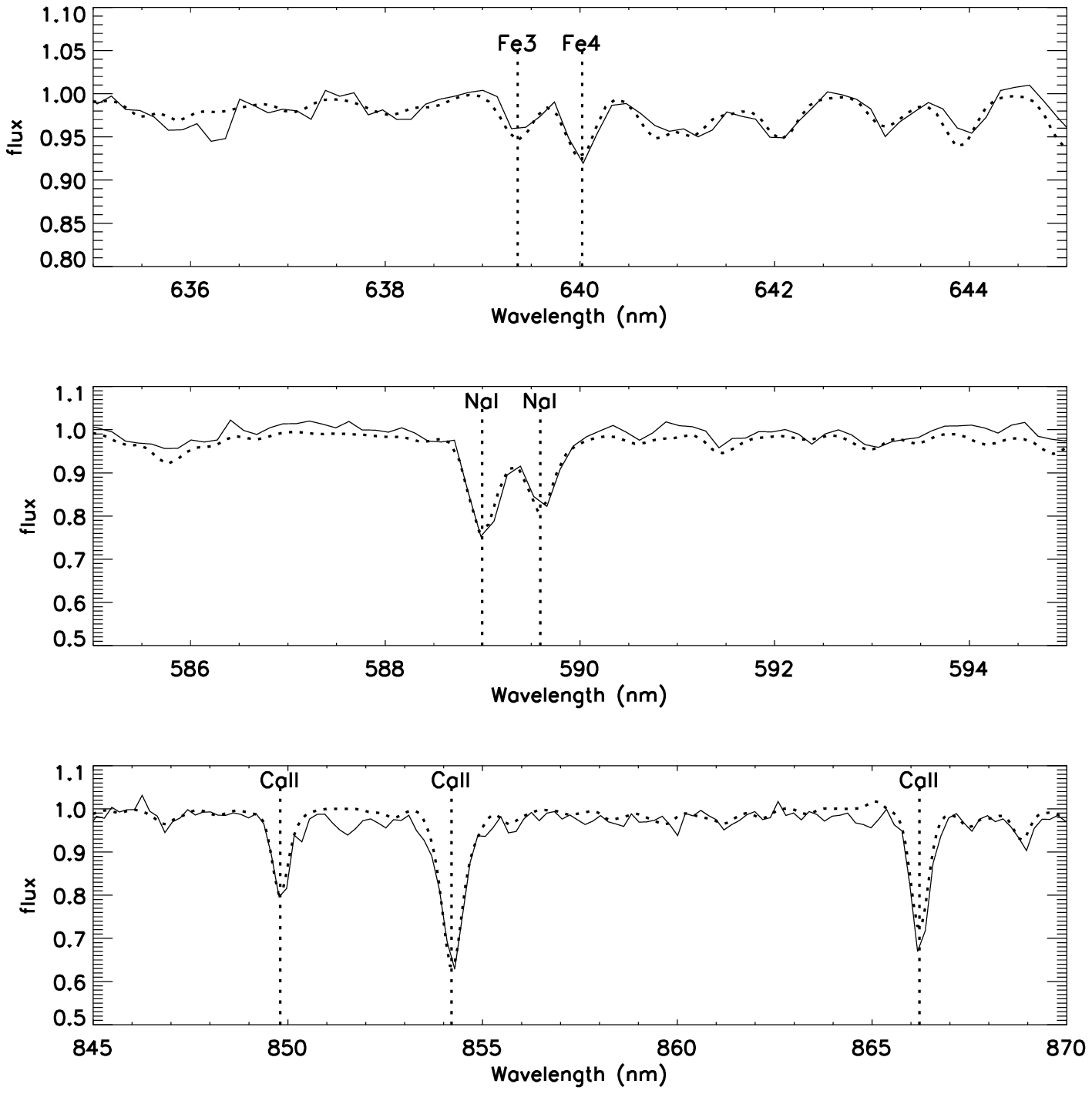}
   \caption{The comparison of the strength of spectral lines. The solid line is the object spectrum and the dotted line is the solar spectrum}
\end{figure}
\clearpage
\begin{figure}
   \vspace{1mm}
   \center
   \hspace{1mm}\includegraphics[width=84mm]{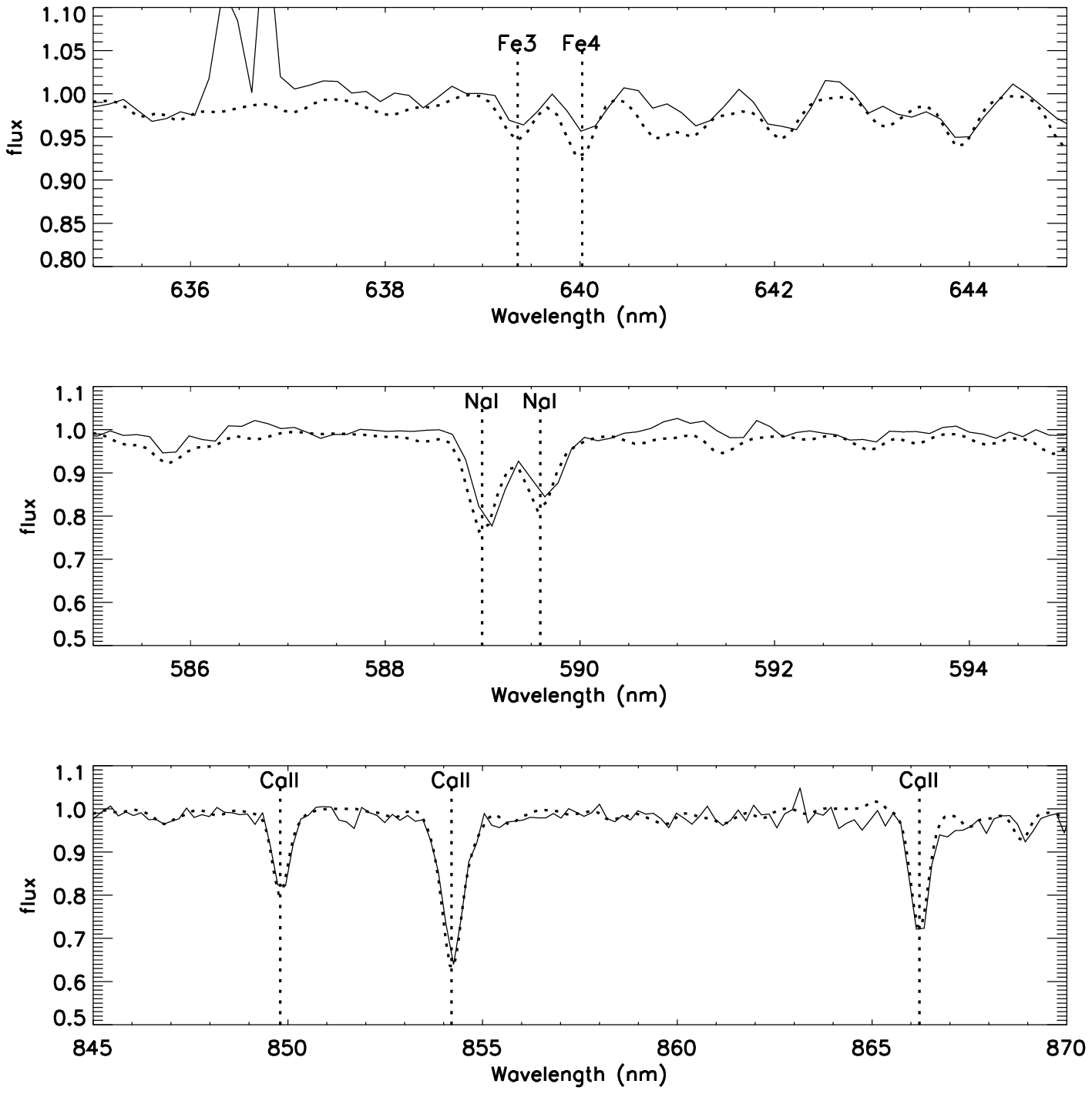}
   \caption{The comparison of the strength of spectral lines. The solid line is the object spectrum and the dotted line is the solar spectrum}
\end{figure}

\begin{figure}
   \vspace{1mm}
   \center
   \includegraphics[width=140mm]{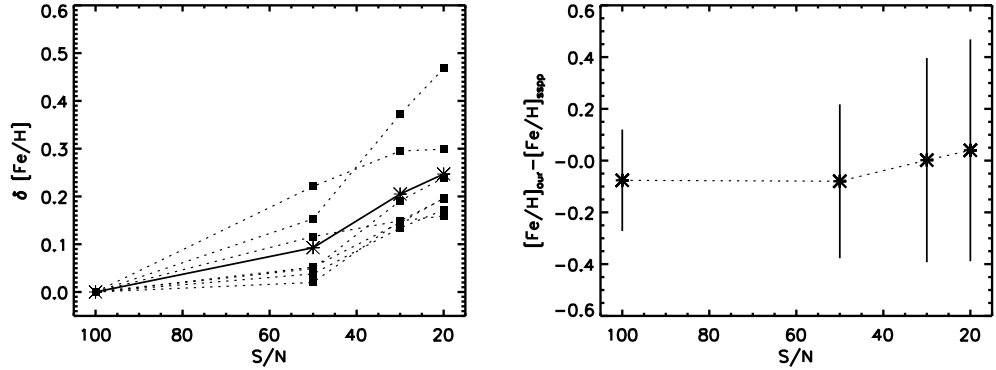}
   \caption{Left:[Fe/H] change vs. S/N for seven stars. The original spectra of these seven stars were extracted from DR7 with S/N $\sim$ 100. By introducing Gaussian noise in original spectra, we degraded them to S/N of 50, 30, and 20, respectively. Filled squares are [Fe/H] changes of these seven stars with different S/N, while asterisks represent average [Fe/H] change for the given S/N. Right: [Fe/H] differences between our calibration results and those from SSPP vs. S/N.  Asterisks represent average [Fe/H] differences and vertical lines represent the scatter of [Fe/H] differences at the given S/N. }
\end{figure}

\subsection{The effect of S/N}
 To investigate the S/N effect on our calibration, we selected seven spectra from DR7 with S/N (r band) $\sim$ 100. By introducing Gaussian noise in these spectra, we degraded them to S/N of 50, 30, and 20, respectively. Then, normalization and EW measurement were carried out to all spectra. Finally, the metallicity were derived with our above calibration. The left panel of Fig. 14 presents the S/N effect on metallicity change. Filled squares represent [Fe/H] changes for these seven stars with different S/N. Asterisks represent the average [Fe/H] changes for the given S/N. It can be seen that [Fe/H] changes about 0.22 dex from S/N = 100 to 20.  [Fe/H] differences between our calibration results and those of SSPP vs. S/N are shown in right panel of Fig. 14. Asterisks are average differences and vertical lines represent scatters for the given S/N for these seven stars. Average [Fe/H] differences nearly keep the same while the scatter will be smaller than 0.4 dex when S/N is higher than 30.  To sum up, our metallicity determination is quite robust to reductions in S/N.
\section{Conclusions}
   For solar type stars, although template matching can derive
reliable results with a suitable wavelength range, it is very difficult to determine the most appropriate wavelength range for matching. 

We selected five FeI lines from the red part of the R$\sim$2000 resolution spectra. These lines, which have a good profile, are not seriously blended and could be detectable with [Fe/H] $>$ -0.8. At the beginning, the metallicity calibrations are set up only through the EW and the scatters are from 0.14 to 0.20 dex. The dispersion becomes small after adding the temperature into the calibrations. Since the temperature is usually unknown in the spectra analysis, it may be good to replace the temperature term with the color index. In view of this, several metallicity calibrations are constructed by the EW of FeI lines and colors based on the 90 solar type stars. The dispersion of all the calibrations is smaller than 0.21 dex. Among the five FeI lines, FeI2 and FeI4 have contributed the better calibrations (Equation 5-6) which have smaller scatters (0.13 dex, 0.15 dex).
\begin{eqnarray}
\rm[Fe/H]=-0.290+8.144*\rm{EW(FeI2)}-1.342*(B-V),~~~ 0.006<EW(FeI2)<0.158
\end{eqnarray}
\begin{eqnarray}
\rm[Fe/H]=-0.321+6.622*\rm{EW(FeI4)}-1.643*(B-V),~~~ 0.029<EW(FeI4)<0.224
\end{eqnarray}

Moreover, we use the calibration from the EW of FeI4 and the B-V to estimate [Fe/H] of the solar type stars in DR7.
After comparing with the value from SSPP, our method gives a good consistency for S/N larger than 50. In
addition, we analyze the stars for which [Fe/H] $\geq0$  by the
spectral lines comparison  and found that 84\% of them are reliable. Usually, [Na/Fe]=0 $\&$ [Ca/Fe]=0 for most stars with [Fe/H] $>$ -0.4 in the solar neighborhood. In view of this, Na and Ca lines are stronger in Fe-rich stars. So this provides a new formula to estimate [Fe/H] with the red band and
presents a reliable way to identify metal rich stars.

\normalem
\begin{acknowledgements}
This work is supported by the National Natural Science
Foundation of China under grant Nos. 10673015, 10821061, 10973021, 11078019 and 11073026, the National Basic Research Program of China (973 program) No.
2007CB815103/815403, the Academy program No. 2006AA01A120 and the
Youth Foundation of the National Astronomical Observatories of China. Many thanks to James Wicker for his help revising English grama of this paper.
\end{acknowledgements}

\label{lastpage}

\end{document}